\documentclass[12pt,english]{article}
\usepackage[T1]{fontenc}
\usepackage[latin2]{inputenc}
\usepackage[letterpaper]{geometry}
\usepackage{textcomp}
\usepackage{amsmath}
\usepackage{amssymb}
\usepackage{graphicx}
\usepackage{setspace}
\makeatletter

\newcommand{\noun}[1]{\textsc{#1}}
\newcommand{\lyxmathsym}[1]{\ifmmode\begingroup\def\b@ld{bold}
  \text{\ifx\math@version\b@ld\bfseries\fi#1}\endgroup\else#1\fi}

\providecommand{\tabularnewline}{\\}

\usepackage{cite}

\makeatother

\usepackage{babel}
\begin{document}

\title{\textbf{Molecular vibrational trapping revisited: a case study with
D2+ }}

\maketitle
\begin{center}
{\large Péter Badankó (1), Gábor J. Halász (2) and Ágnes Vibók (1,3)}\\
(1) Department of Theoretical Physics, University of Debrecen, \\
H-4010 Debrecen, PO Box 5, Hungary\\
(2) Department of Information Technology, University of Debrecen,
\\
H-4010 Debrecen, PO Box 12, Hungary\\
(3) ELI-ALPS, ELI-HU Non-Profit Ltd,\\
H-6720 Szeged, Dugonics ter 13, Hungary
\par\end{center}
\begin{abstract}
The present theoretical study is concerned with the vibrational trapping
or bond hardening, which is a well-known phenomenon predicted by a
dressed state representation of small molecules like $H_{2}^{+}$
and $D_{2}^{+}$in an intense laser field. This phenomenon is associated
with a condition where the energy of the light induced, vibrational
level coincides with one of the vibrational levels on the field-free
potential curve, which at the same time maximizes the wave function
overlap between these two levels. One-dimensional numerical simulations
were performed to investigate this phenomenon in a more quantitative
way than has been done previously by calculating the photodissociation
probability of $D_{2}^{+}$ for a wide range of photon energy. The
obtained results undoubtedly show that the nodal structure of the
field-free vibrational wave functions plays a decisive role in the
vibrational trapping, in addition to the current understanding of
this phenomenon. 

\end{abstract}

\section{Introduction}

Understanding the behavior of atoms and molecules in a strong laser
field is an intensively studied area and there are numerous experimental
and theoretical investigations which have uncovered and explored several
new phenomena of light-matter interactions including high harmonic
generation, above threshold ionization or dissociation \cite{HHG,ATI,ATI3,ATI4}.
The bond softening and the bond hardening effects \cite{Bandrauk1,Bandrauk2,Bandrauk3,Bandrauk4,Bandrauk5,Bucksbaum1,Bucksbaum2,Bucksbaum3,Bucksbaum4,Giusti1,Giusti2,Post1,Post2,Esry1,Atabek,Atabek2,Whelan,Vrakking,Dorner,Natan}
are similar phenomena which are often visualized during the photodissociation
or photofragmentation processes of diatomic or polyatomic molecules.
The mechanism of bond softening is well known. It was first discovered
experimentally in the dissociation spectra of the $H_{2}^{+}$ and
$D_{2}^{+}$ ions by Bucksbaum and coworkers\cite{Bucksbaum1,Bucksbaum2,Bucksbaum3}
and can easily be understood by the illustrative Floquet's picture
\cite{chu1,chu2} or dressed state representation, which is often
used to explain various strong field physics phenomena. The Floquet
Hamiltonian for the net absorption of one photon can be represented
by a 2$\times$2 matrix, which explicitly includes the light-matter
interaction. The change of the nuclear dynamics, due to the laser
field, can be visualized as arising from ``light-induced'' or ``adiabatic''
molecular potentials (LIP) \cite{Bandrauk6}. The molecular potential
deforms due to strong radiative coupling and results in a strongly
enhanced dissociation rate. The adiabatic curves show that as the
laser intensity increases, the dissociation barrier moves to lower
vibrational levels, leading to a noticeable growth of the dissociation
probability. Bond hardening, which is often called molecular stabilization
or vibrational trapping is the opposite effect with the same origins.
Despite numerous experimental and theoretical works that have been
performed in order to gain a better qualitative understanding of the
mechanism behind this phenomenon \cite{Bandrauk1,Bandrauk2,Bandrauk3,Bandrauk4,Bandrauk5,Bucksbaum1,Bucksbaum2,Bucksbaum3,Bucksbaum4,Giusti1,Giusti2,Frasinski1,Sandig1,Nimrod,Moiseyev-Pawlak},
there still remained some unresolved issues. 

Recent efforts have been invested in studying the nature of the light-induced
conical intersections (LICIs) \cite{Milan,Nimrod} which was first
discussed for diatomics by applying Floquet representation. In a diatomic
molecule, which has only one nuclear vibrational coordinate, it is
not possible for two electronic states of the same symmetry to form
a conical intersection (CI). However, this former statement is only
true in field-free space. Conical intersections can be formed, even
in diatomic molecules, if an additional degree of freedom e.g. rotation,
is associated with the system exposed to strong laser fields. In this
situation, the interaction of the transition dipole moment of the
molecule with the external electric field leads to an effective torque
toward the polarization direction of the light. By considering the
rotational coordinate in the description as a dynamical variable,
the change of nuclear dynamics due to the external light can be considered
as arising from a LICI. The positions of these LICIs are determined
by the laser frequency while the laser intensity controls the strength
of the nonadiabatic coupling. Several theoretical and experimental
works have demonstrated that the LICIs have strong impact on the spectroscopic
and dynamical properties of molecules \cite{Gabia,Gabib,Gabic,Gabid,Gabie,Gabi1,Gabi2,Gabi3,Gabi4,Gabi5,Moiseyev-Pawlak,Buks2,Buks3,Buks4,Sola1,Sola2,Philip}. 

There are many topical studies investigating the dissociation process
of the $D_{2}^{+}$ molecule within the LICI framework \cite{Gabi1,Gabie,Gabi2,Gabi3,Gabi4}.
In these works it was found that using relatively small intensities
($1\times10^{11}$ to $1\times10^{12}$ W/cm$^{2}$) and propagating
the dynamics of the $D_{2}^{+}$ results in very small (practically
zero) rates for the dissociation probabilities of the photofragments
\cite{Gabi2} for certain vibrational eigenstates and photon energies.
The dissociation yield still remains relatively small even at the
highest studied intensity ($1\times10^{14}$ W/cm$^{2}$). The obtained
results are almost independent of the origins of the initial nuclear
wave packet, which can start either from a vibrational eigenstate
or a superposition of eigenstates, the Franck - Condon (FC) distribution,
of the $D_{2}^{+}$.

Intensity dependence at fixed frequency regarding the vibrational
trapping phenomenon in the $H_{2}^{+}$ and $D_{2}^{+}$ ions was
intensively theoretically studied by Bandrauk and coworkers\cite{Bandrauk1,Bandrauk2,Bandrauk3,Bandrauk4,Bandrauk6}.
They highlighted that laser-induced avoided crossings provide a very
important source of molecular stabilization and hence suppress dissociation
at high intensities. They also found that when the diabatic (field-free)
and the adiabatic (light-induced) vibrational levels coincide, the
resulting resonances can minimize the photodissociation probabilities.
Increasing the field intensities, trapping of the initial state into
stable adiabatic states occurs due to the decreasing nonadiabatic
couplings between the light-induced states. Their work resulted in
a qualitatively correct picture of the vibrational trapping phenomenon
mechanism. 

In this article we investigate why, in certain situations, there is
virtually no dissociation rate of the photofragments arising from
the dissociation process of the $D_{2}^{+}$ molecule. This paper
expands previous investigations by trying to find a more accurate
and quantitative connection between physical quantities including
the coincidence of the diabatic and adiabatic vibrational levels,
maximum overlaps between the diabatic and adiabatic wave functions
and minima of the photodissociation probabilities. This is realized
by calculating the difference between the diabatic and adiabatic vibrational
energies, the overlap between the diabatic and adiabatic wave functions
and the dissociation yield of the molecular photofragments over a
wide range of photon energies (from $\omega_{1}=24.799$ eV, $\lambda_{1}=50$
nm to $\omega_{2}=3.100$ eV, $\lambda_{2}=400$ nm). 

As was discussed in \cite{Gabi2,Gabi3}, the bond hardening effect
is highly dependent on the behavior of the system in low intensity
field even if the actual applied intensity does not fall into this
region. This is related to the fact that --not considering few cycle
pulses-- larger intensity pulses are built up gradually. Consequently,
at the beginning of the pulse the role of the different adiabatic
levels is determined by the low intensity behavior of the system.
For this reason we use $1\times10^{11}$ W/cm$^{2}$ intensity in
our simulations or even the zero intensity limit to determine the
different features of the vibrational eigenstates of the adiabatic
upper state. In the simulations we included only the two lowest lying
electronic states of the $D_{2}^{+}$molecule. This decreases the
validity of the results for the higher energy region -- especially
for the $\lambda\lesssim100nm$ -- where the Rydberg states of the
ion and even the doubly-ionized repulsive Coulomb states can play
an essential role in the real processes. Fortunately, this inaccuracy
in the description of the physical system does not invalidate our
findings related to the bond hardening effect, which is widely studied
within the two state approximation. One can consider that some of
our results are valid only for a model system inspired by the $D_{2}^{+}$
molecular ion.

\section{Theory and methods }

\begin{figure}[p]
\includegraphics[width=0.5\textwidth]{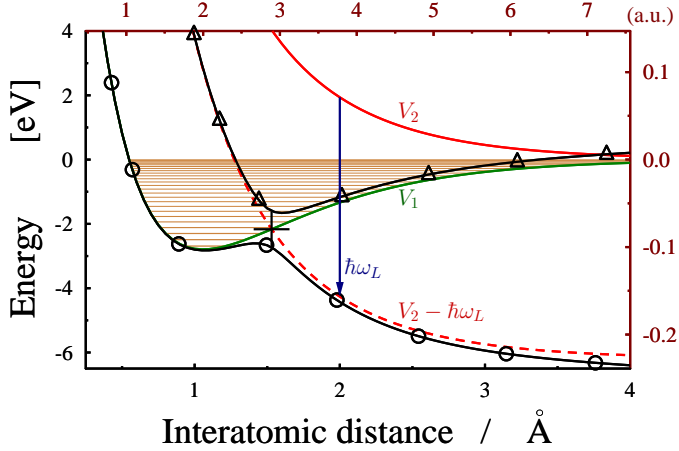}

\caption{A cut through the potential energy surface of the $\mathrm{D}_{2}^{+}$
molecule as a function of inter atomic separation. Diabatic energies
of the ground $\left(V_{1}\right)$ and the first excited $\left(V_{2}\right)$
states are displayed with solid green and red lines, respectively.
The field dressed excited state ($V_{2}-\hbar\omega_{L}$ ; dashed
red line) forms a light induced conical intersection (LICI) with the
ground state. For the case of a laser frequency $\omega_{L}=6.199\,\, eV$
and field intensity of $3\times10^{13}\frac{W}{cm^{2}}$ a cut through
the adiabatic surfaces at $\theta=0$ (parallel to the field) is also
shown by solid black lines marked with circles ($V_{lower}$) and
triangles ($V_{upper}$). The position of the LICI is indicated with
a cross. }

\label{potential}
\end{figure}

Figure \ref{potential} shows the potential energy curves for $D_{2}^{+}$
in dressed state representation. The electronic ground ($V_{1}(R)=1s\sigma_{g}$)
and the first excited ($V_{2}(R)=2p\sigma_{u}$) eigenstates are considered
as diabatic potentials and together, with the kinetic energy, form
the field free Hamiltonian

\begin{equation}
H_{diabatic}^{f-free}=\left(\begin{array}{cc}
-\frac{1}{2\mu}\frac{\partial^{2}}{\partial R^{2}}+V_{1}(R) & \;0\\
0 & \;-\frac{1}{2\mu}\frac{\partial^{2}}{\partial R^{2}}+V_{2}(R)
\end{array}\right)\label{eq:1}
\end{equation}
where $R$ is the vibrational coordinate and $\mu$ is the reduced
mass. The ion is excited by a resonant laser pulse from the $V_{1}(R)$
ground state to the repulsive $V_{2}(R)$ state. An electronic transition
occurs due to the nonvanishing transition dipole moment and the corresponding
$2\times2$ field dressed Hamiltonian matrix reads

\begin{equation}
H_{dressed}=\left(\begin{array}{cc}
-\frac{1}{2\mu}\frac{\partial^{2}}{\partial R^{2}}+V_{1}(R) & (\epsilon_{0}/2)d(R)\cos\theta\\
(\epsilon_{0}/2)d(R)\cos\theta & -\frac{1}{2\mu}\frac{\partial^{2}}{\partial R^{2}}+V_{2}(R)-\hbar\omega
\end{array}\right).\label{eq:2}
\end{equation}
The off-diagonal elements of eq. (\ref{eq:2}) represent the radiative
couplings, where $\epsilon_{0}$ is the laser field amplitude, $\omega$
is the laser frequency which couples the two electronic states, $d(R)$
is the transition dipole and $\theta$ is the angle between the polarization
direction of the light and the direction of the molecular axes. The
potential energies of $V_{1}(R)$ and $V_{2}(R)$ and the transition
dipole moment were sourced from refs. \cite{Pot} and \cite{dipol}.
The exact, time-dependent (TD) form of eq. (\ref{eq:2}), which is
used extensively in the time-dependent calculations is also given
here

\begin{equation}
H_{TD}=\left(\begin{array}{cc}
-\frac{1}{2\mu}\frac{\partial^{2}}{\partial R^{2}}+V_{1}(R) & \epsilon_{0}f(t)d(R)\cos\theta\cos\omega t\\
\epsilon_{0}f(t)d(R)\cos\theta\cos\omega t & -\frac{1}{2\mu}\frac{\partial^{2}}{\partial R^{2}}+V_{2}(R)
\end{array}\right)\label{eq:3}
\end{equation}
where $f(t)$ is the envelope function of the laser pulse. 

\begin{figure}[p]
\includegraphics[width=0.5\textwidth]{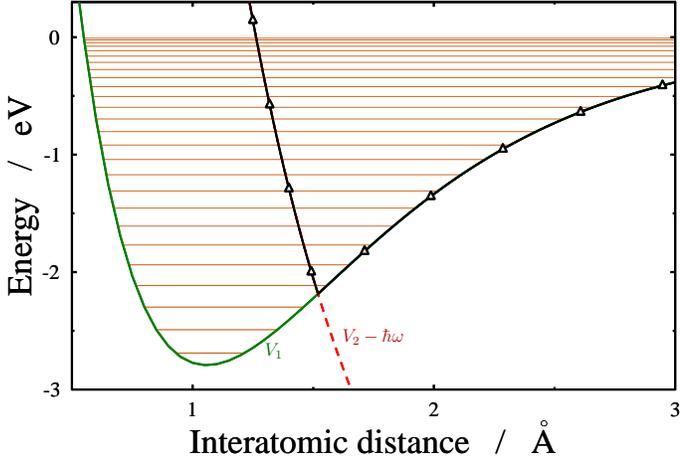}

\caption{Diabatic and adiabatic cut of the potential energy surfaces along
the inter atomic separation. Diabatic energies of the ground ($V_{1}$;
solid green) and the field dressed excited ($V_{2}-\hbar\omega_{L}$
; dashed red line) states are displayed. The adiabatic upper curve
--calculated by using eq. (\ref{eq:4}) for the weak field limit--
is denoted by solid black line marked with triangles.The different
horizontal lines provide the appropriate diabatic (in orange) vibrational
energy levels. }

\label{Intensity&wavelength}
\end{figure}

A convenient and approximate form for the adiabatic potential is assumed.
The energy levels and the eigenfunctions of the upper adiabatic potential
can easily be calculated for weak fields from a model potential (Figure
\ref{Intensity&wavelength}): 

\begin{align}
H_{adiabatic}^{upper} & =T_{kin}+V_{model}\label{eq:4}\\
 & =-\frac{1}{2\mu}\frac{\partial^{2}}{\partial R^{2}}+(V_{2}-\hbar\omega)\cdot\Theta(R_{CR}(\lambda)-R)+V_{1}\cdot\Theta(R-R_{CR}(\lambda))\nonumber 
\end{align}
where $\Theta$ is the Heaviside step function, $\lambda$ is the
wavelength and $R_{CR}$ is the crossing point between the ground
$V_{1}(R)$ and the field dressed ($V_{2}(R)-\hbar\omega$) excited
states. This model potential in eq. (\ref{eq:4}) can be converted
into the upper adiabatic potential at zero-field limit. It is now
possible to calculate\textit{ }the vibrational energy levels ($E_{\nu}$)
and wave functions ($\psi_{\nu}$) belonging to the different vibrational
eigenstates of the diabatic potential in eq. (\ref{eq:1}) and the
eigenvalues ($\mathcal{E}_{\nu'}$) and eigenfunctions ($\varphi_{\nu'}$)
of the upper adiabatic potential in eq. (\ref{eq:4}). A more detailed
analysis requires new quantities derived from different combinations
of the previously determined basic quantities. These are the difference
of the adiabatic and diabatic eigenvalues

\begin{equation}
\Delta E_{\nu,\nu'}(\lambda)=\mathcal{E}_{\nu'}-E_{\nu};\label{eq:5}
\end{equation}
the adiabatic and diabatic eigenfunctions overlap, 
\begin{equation}
S_{\nu,\nu'}(\lambda)=<\varphi_{\nu'}(\lambda,R)|\psi_{\nu}(R)>=\int\limits _{-\infty}^{\infty}\varphi_{\nu'}^{*}(\lambda,R)\psi_{\nu}(R)dR\label{eq:6}
\end{equation}
and the total dissociation probability:

\begin{equation}
P_{diss}=\int\limits _{0}^{\infty}dt<\psi(t)|W|\psi(t)>.\label{eq:7}
\end{equation}
The $\Delta E_{\nu,\nu'}(\lambda)$ and $S_{\nu,\nu'}(\lambda)$ were
determined with a spacing of $\Delta\lambda=1.0$ in the interval
of ($50$ nm -- $400$ nm), $-iW$ is the complex absorbing potential
(CAP) applied at the last $5$ a.u. of the grid related to the vibrational
degree of freedom ($W=0.00005\cdot(r\lyxmathsym{\textminus}70)^{3}$,
if $r>70$ a.u. on the $1s\sigma_{g}$ surface and $W=0.00236\cdot(r\lyxmathsym{\textminus}75)^{3}$,
if $r>75$ a.u. on the $2p\sigma_{u}$ surface.

The multi-configuration time-dependent Hartree (MCTDH) method \cite{Dieter1,Dieter2,Dieter3,Dieter4,Dieter5}
is used to solve the time-independent and time-dependent Schrödinger
equations in conjunction with the eigenvalue and eigenfunction problems
in the diabatic/adiabatic frameworks, (eqs. (\ref{eq:1}) and (\ref{eq:4}))
and the dissociation dynamics (eq. (\ref{eq:3})). Fast Fourier transformation-discrete
variable representation (FFT-DVR) \cite{Kosloff} is used to characterize
the vibrational degree of freedom, with $N_{R}$ basis elements for
the internuclear separation distributed within the range from $0.1$
a.u. to $10.05$ a.u. or $80$ a.u. in the eigenstates or the dissociation
yield calculations, respectively. These primitive basis sets ($\chi$)
are used to represent the wave function
\begin{equation}
\psi(R,t)=\sum\limits _{l=1}^{N_{R}}c_{j_{R}l}^{(R)}(t)\chi_{l}^{(R)}(R).\label{eq:8}
\end{equation}
In the actual calculations, $N_{R}=256$ is used for the eigenstates
and $N_{R}=2048$ for the dynamical calculations, respectively.

\section{Results and discussion}

Our present studies intensively probe the effect of molecular stabilization
using accurate numerical calculations, performed in a wide range of
energies. Detailed analysis is performed using a one-dimensional (1D)
nuclear dynamical simulation and the initial nuclear wave packet is
assumed to be in one of the vibrational eigenstates ($\nu=0,\,1,\,2,\,3,\,4,\,5,\,6,\,7,\,8,\,9$)
or Franck-Condon distribution of the vibrational states of the ion.
The initial orientation of the molecules is assumed to be parallel
to the external field in all cases. The dissociation rates for isotropic
initial distribution can be approximated with large accuracy by dividing
the results by 3 as this study is limited to low intensities. The
energy interval ( $\omega_{1}=24.799$ eV, $\lambda_{1}=50$ nm to
$\omega_{2}=3.100$ eV, $\lambda_{2}=400$ nm) is chosen, so that
it contains all the values belonging to the positions of the near
zero dissociation probability of the studied vibrational levels. A
linearly polarized Gaussian laser pulse, with intensity, $I_{0}=10^{11}W/cm^{2}$
and a pulse duration in full width at half-maximum (FWHM) $t_{pulse}=30$
fs is used throughout the calculations.

Three physical properties, the difference of the diabatic and adiabatic
eigenvalues ($\Delta E(\lambda)$), the diabatic and adiabatic eigenfunctions'
overlap ($S(\lambda)$) and the dissociation probability ($P_{diss}$)
play an important role in the forthcoming analysis.

\subsection{Dissociation probability, wave function overlap and energy difference}

The dissociation probability, eq. (\ref{eq:7}), is calculated in
the chosen energy interval by maintaining a fixed light intensity.
Wavelengths which result in a virtually zero dissociation probability
from a particular vibrational energy level are sought, like in previous
investigations \cite{Gabi2}. Calculations have been performed for
all chosen vibrational eigenstate, $\nu=0,\,1,\,\cdots,\,9$, and
the previously determined energy interval ensured that for each eigenstates
all the wavelengths which belong to a minimum value of the dissociation
probability were assigned. For example, nine different values for
the wavelength were found for the $\nu=9$ vibrational level . 

\begin{figure}[p]
\includegraphics[width=0.5\textwidth]{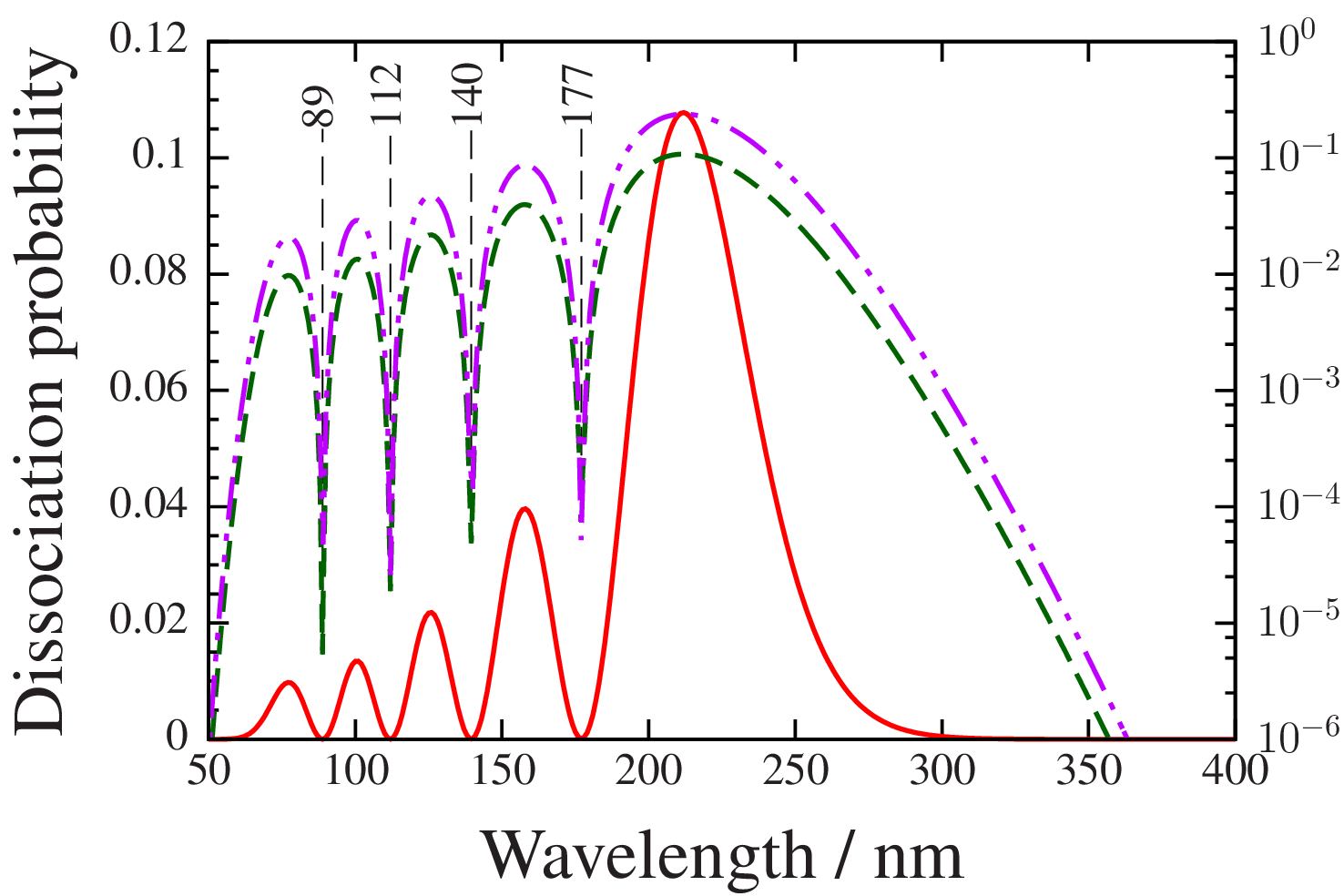}

\caption{Total dissociation probabilities as a function of laser wavelength
for the $\nu=4$ vibrational eigenstate. The applied energy/wavelength
interval and the intensity are from $\hbar\omega_{1}=24.798\, eV$,
$\lambda_{1}=50\, nm$ to $\hbar\omega_{2}=3.0998\, eV$, $\lambda_{2}=400\, nm$
and $I=1\times10^{11}\frac{W}{cm^{2}}$, respectively. The total dissociation
probability from the $\nu=4$ vibrational eigenstate is displayed
by the solid (linear scale on the left side) and dashed (logarithmic
scale on the right side) curves. The dashed--dotted curve displays
(logarithmic scale on the right side) the differential dissociation
rate with kinetic energy of fragments being $E_{\nu}+\hbar\omega$
when the simulation was started from Franck--Condon distribution.}

\label{dissociation}
\end{figure}

Figure \ref{dissociation} shows the results for the $\nu=4$ vibrational
level for the case when the initial nuclear wave packet starts from
(i) a vibrational eigenstate or from (ii) a superposition of eigenstates,
the Franck - Condon, (FC) distribution, of the $D_{2}^{+}$. The dissociation
rate has four minima at the $\lambda_{D}(4,3)=88.8$ nm, $\lambda_{D}(4,2)=112$
nm, $\lambda_{D}(4,1)=139.5$ nm and $\lambda_{D}(4,0)=177.1$ nm
wavelengths. Simulations in which the wave packets start from a vibrational
eigenstate use pulses with a $30$ fs duration centered around $0$
fs and the total dissociation probabilities are calculated. In the
superposition situation, pulses with a $30$ fs duration are applied.
These pulses are centered around $34$ fs during which time approximately
one and a half vibration cycle takes place on the diabatic lower surface
$V_{1}$ \cite{Gabi1}. The yield of total dissociation probability
does not provide any information about the dissociation rate resulting
from a particular vibrational state because all of the vibrational
eigenstates are included in the initial wave packet. Thus, attention
turned to the kinetic energy (KER) release spectra and the dissociation
rate at the $E_{\nu}+\hbar\omega$ energy was studied.

Results for the dissociation yield obtained by assuming FC distribution
of the nuclear wave packet are surprisingly similar to those obtained
from initiating the dynamics from one of the vibrational eigenstates
of the Hamiltonian. This similarity is related to the fact that, on
one hand the applied laser pulse is long enough to resolve properly
the vibrational energy eigenstates and on the other hand its intensity
is in the linear respond regime. These two features allow us to conclude
that the structure of the initial wave packet does not significantly
affect the underlying physical mechanism in the bond hardening effect.
(Not displayed results for other vibrational states clearly confirm
this statement, as well.) For the sake of simpler analysis initial
wave packets assuming vibrational eigenstate form are used in the
further calculations.

\begin{figure}[p]
\includegraphics[width=1\textwidth]{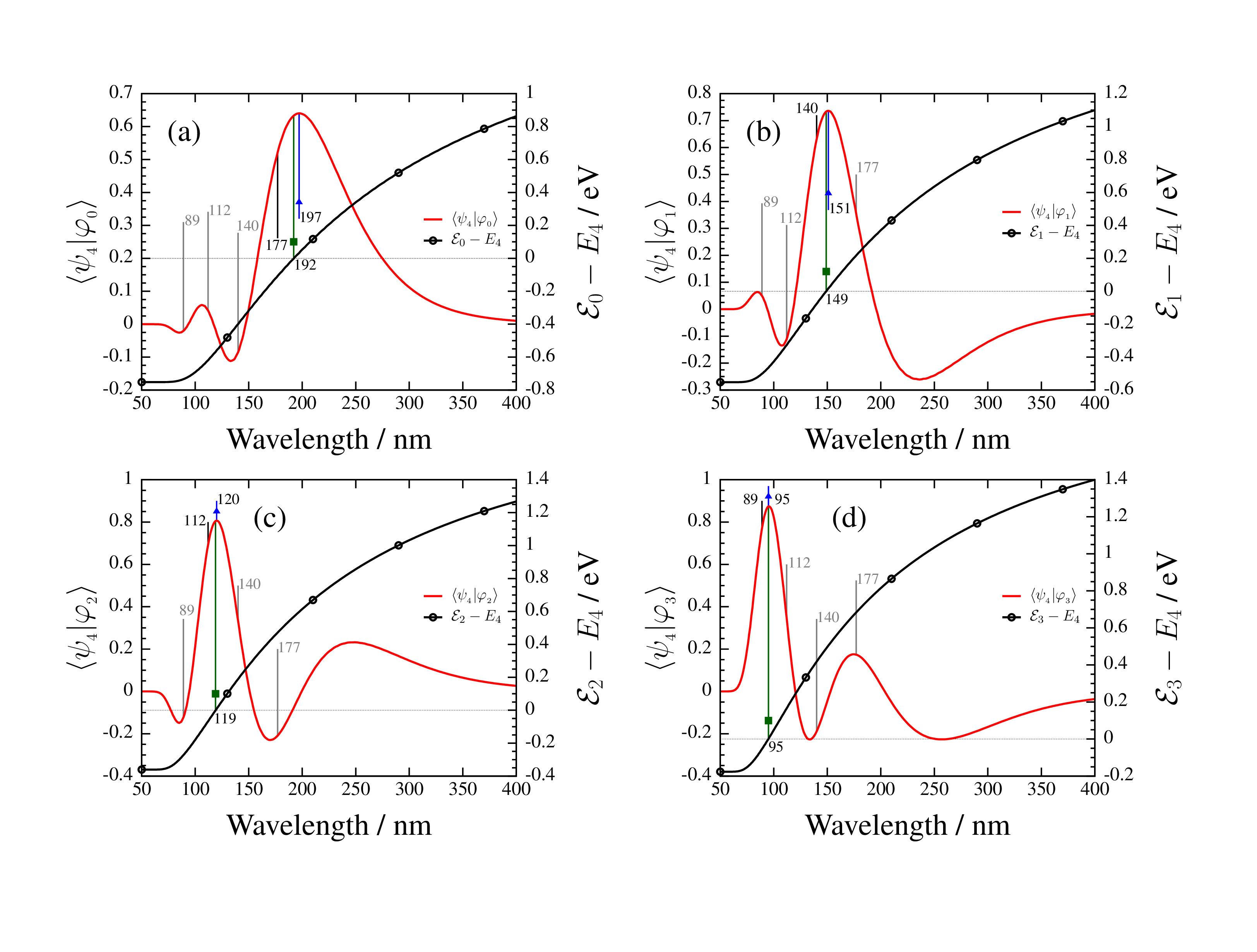}

\caption{Overlap $S_{\nu,\nu'}(\lambda)$ between the diabatic ($\psi_{\nu}$;
$\nu=4$) and adiabatic ($\varphi_{\nu'}$) eigenfunctions and the
difference of the diabatic and adiabatic eigenvalues ($\Delta E_{\nu,\nu'}(\lambda)=\mathcal{E}_{\nu'}-E_{\nu}$;
$\nu=4$) as a function of wavelength. The value of $S_{\nu,\nu'}(\lambda)$
(solid curve) is given by the scale on the left side, while the value
of $\Delta E_{\nu,\nu'}(\lambda)$ (dotted curve) is given by the
scale on the right side. Bars denote the wavelengths for which the
total dissociation probability is minimal. Bars with triangle and
bars with square present the $\lambda_{O}(\nu,\nu')$ and $\lambda_{E}(\nu,\nu')$
wavelengths corresponding to the maximum value of $S_{\nu,\nu'}(\lambda)$
and the zero value of $\Delta E_{\nu,\nu'}(\lambda)$, respectively.
In panel (a) $(\nu'=0),$ in panel (b) $(\nu'=1),$ in panel (c) $(\nu'=2)$
and in panel (d): $(\nu'=3)$. }
\label{overlap}
\end{figure}

The next task was to obtain the diabatic and adiabatic eigenfunctions'
overlap which was calculated using eq (\ref{eq:6}). For each diabatic
vibrational level ($\nu$) $\nu'=0,\,1,\,\cdots,\nu-1$, adiabatic
eigenstates and correspondingly $\nu$ different overlap functions
exist. Figure \ref{overlap} shows these functions for the $\nu=4$
vibrational level. The $4$ different panels belong to the corresponding
values of $\nu'$ from 0 to $\nu-1,$ and the obtained curves are
scaled on the left side of the panels.

The difference of the adiabatic and diabatic eigenvalues eq (\ref{eq:5}),
is determined and $\nu$ energy difference curves are obtained for
each value of $\nu$. Figure \ref{overlap} shows these curves which
are scaled on the right side of the panels. It can be expected, based
on previous theoretical studies \cite{Bandrauk4,Bandrauk5,Bandrauk6}
that, for a given pair of diabatic and adiabatic eigenstates with
the feature of $\nu>\nu'$, both the maxima of the overlap between
the diabatic and adiabatic vibrational wave functions at $\lambda=\lambda_{O}(\nu,\nu')$
and the matching of the diabatic and adiabatic eigenvalues at $\lambda=\lambda_{E}(\nu,\nu')$
are closely related to the minimal dissociation yield or bond hardening
situation, which takes place at special photon energies $\lambda=\lambda_{D}(\nu,\nu')$.
These special values of the wavelength are presented in the first
four columns of Table \ref{Table1} for all the studied vibration
levels ($\nu=0,\,1,\,\cdots,\,9$). If the values of these three wavelengths
($\lambda_{D}$, $\lambda_{O}$ and $\lambda_{E}$) are close to each
other for a certain pair of vibrational frequency ($\nu,\,\nu'$),
the values of the $\frac{\lambda_{O}}{\lambda_{D}}$ and $\frac{\lambda_{E}}{\lambda_{D}}$
ratios are also close to 1. However, figure \ref{ratio} shows that
this is not the case as the $\frac{\lambda_{O}}{\lambda_{D}}$ and
$\frac{\lambda_{E}}{\lambda_{D}}$ wavelength rates systematically
deviate from 1 within a range of $5\%$--$15\%$ for all cases. The
significantly large differences between the wavelengths suggest that
the physics behind the trapping effect may not be described adequately
by studying the overlap of the wave functions or energy differences
and it is expected, that some other factors may play a role.

\subsection{The role of the nodal structure}

\begin{figure}[p]
\includegraphics[width=0.5\textwidth]{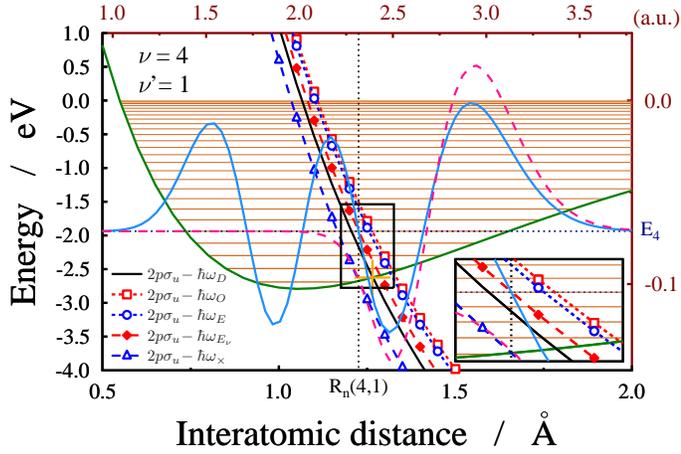}

\caption{The $\psi_{\nu}(\nu=4)$ diabatic eigenstate (cyan (light gray) line)
and $\varphi_{\nu^{'}}(\nu'=1)$ adiabatic eigenstate (dashed pink
curve) for the case of maximal overlap ($\lambda_{O}(4,1)$) with
the diabatic $\nu=4$ eigenstate. The zero line for the wave functions
is placed to the energy level of the $\nu=4$ vibrational state. The
$V_{1}$ diabatic and $V_{2}$ potential energies with five field-dressed
excited states are also shown. The solid black line denotes the shift
by a laser field with the wavelength $\lambda_{D}(4,1)$ ($\omega_{D}(4,1)$
frequency), for which the dissociation of the $D_{2}^{+}$ is minimal.
Lines marked with empty squares and circles are shifted by the energies
of $\hbar\omega_{O}$ and $\hbar\omega_{E}$, respectively. The line
marked with solid diamonds corresponds to a shift with $\hbar\omega_{E_{4}}$,
so that the energy of the dressed state at the third node ($R_{n}(4,1)$)
of the $\nu_{4}$ diabatic eigenfunction equals to the $E_{4}$ vibrational
eigen energy. The $\hbar\omega_{\times}$ is the photon energy ($\lambda_{\times}(4,1)$
wavelength) by which shifting the $V_{2}$ state (marked with triangles)
a crossing with the ground state potential is formed at the position
of the $R_{n}(4,1)$ node. The inset in the bottom right corner displays
a zoomed version of the central area.}
\label{nodes}
\end{figure}

This section starts with a simple analysis of the adiabatic and diabatic
wave functions overlap. It is obvious that the node positions of adiabatic
wave function should be close to the node positions of the diabatic
function in order to obtain a significantly large overlap. The adiabatic
wave function fades to zero at such internuclear distances where the
adiabatic potential is above the adiabatic energy level. Applying
a photon energy which corresponding to the dissociation minimum results
in the energies of the diabatic and adiabatic states being similar.
In such a case, the disappearance of the adiabatic wave function can
be expected to approximately take place where the adiabatic potential
is equal to the energy of the diabatic eigenstate. Thus, it can be
assumed that for a large overlap the matching of the adiabatic potential
to the diabatic energy level should take place around one of the nodes
of the diabatic eigenfunction. The wavelength $\lambda_{E_{\nu}}(\nu,\nu')$
is where the matching of the adiabatic potential to the energy of
the diabatic eigenstate takes place at the $\left(\nu-\nu'\right)^{th}$
node $\left(R_{n}(\nu,\nu')\right)$ of the diabatic wave function,
i.e., after which there are additional $\nu'$ nodes ( Figure \ref{nodes}):
\begin{equation}
\hbar\omega_{E_{\nu}}=V_{2}\left(R_{n}(\nu,\nu')\right)-E_{\nu}.\label{eq:9}
\end{equation}

There is another possible way to demonstrate the importance of the
nodes of the nuclear wave packet in the process of molecular stabilization.
In the light-induced picture of dissociation, the nonadiabatic coupling
between the two surfaces will be the largest around the crossing of
the ground and the dressed excited state diabatic potentials. Therefore,
a reasonable reduction in the dissociation yield can be expected if
the initial diabatic wave packet has a node at, or close to this crossing.
To check the validity of this approach, the wavelengths for which
the crossing of the potentials takes place at one of the nodes of
the diabatic wave function was determined. These values are denoted
as $\lambda_{\times}\left(\nu,\nu'\right)$ and the corresponding
photon energies can be calculated as
\begin{equation}
\hbar\omega_{\times}=V_{2}\left(R_{n}(\nu,\nu')\right)-V_{1}\left(R_{n}(\nu,\nu')\right).\label{eq:10}
\end{equation}
Table \ref{Table1} shows both of these newly introduced quantities
and the positions of their dependant nodes.

\begin{figure}[p]
\includegraphics[width=0.5\textwidth]{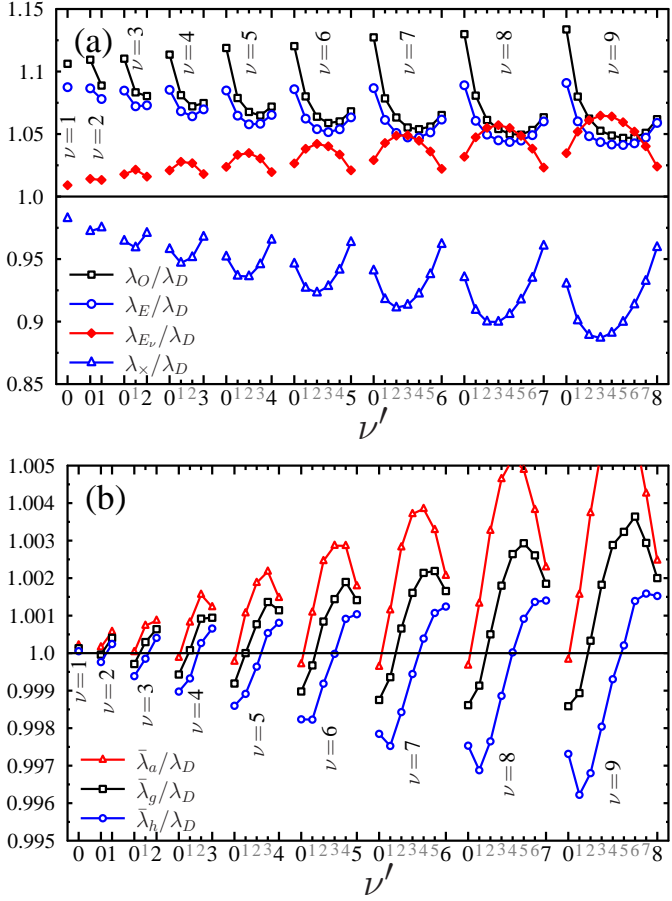}

\caption{Ratios of the different wavelengths with respect to the minimum value
of the dissociation probability $\lambda_{D}$. Panel (a): $\lambda_{O}$
is the wavelength belonging to the maximum value of the wave functions'
overlap, $\lambda_{E}$ is the wavelength where the adiabatic and
diabatic eigenvalues are the same, $\lambda_{E_{\nu}}$ is the wavelength
according to eq. (\ref{eq:9}) and $\lambda_{\times}$ is the wavelength
according to eq. (\ref{eq:10}); Panel (b): $\bar{\lambda}_{a}$,
$\bar{\lambda}_{g}$ and $\bar{\lambda}_{h}$ are the weighted arithmetic,
geometric and harmonic means of $\lambda_{E_{\nu}}$ and $\lambda_{\times}$
as defined in eqs. (\ref{eq:11}--\ref{eq:13}).}
\label{ratio}
\end{figure}

The accuracy of the above constructed approach was checked by calculating
the ratios with $\lambda_{D}$ and the obtained $\frac{\lambda_{\times}}{\lambda_{D}}$
and $\frac{\lambda_{E_{\nu}}}{\lambda_{D}}$ rates are shown in Figure
\ref{ratio}a. For lower lying vibrational states, both these two
new formulas, based on the position of the nodes, provide a better
forecast of the wavelength of the minimal dissociation yield than
the preliminary defined $\lambda_{E}$ or $\lambda_{O}$ values. $\lambda_{\times}$
underestimates $\lambda_{D}$ approximately twice as much as $\lambda_{E_{\nu}}$
overestimates. This observation can be used to construct some additional
semiempirical formulas which might balance these two opposite deviations
in the $\lambda_{\times}$ and $\lambda_{E_{\nu}}$wavelengths (see
in Fig. \ref{ratio}a). In these formulas the $\lambda_{E_{\nu}}$
wavelength always takes part with twice the weight in. Obtained combinations
are: the weighted arithmetic, geometric and harmonic means (arithmetic
mean of the photon energies) of the former two node based approximations
for the position of the minimal dissociation rate:
\begin{align}
\bar{\lambda}_{a} & =(\lambda_{\times}+2\lambda_{E_{\nu}})/3 & \Longleftrightarrow\quad & \bar{\omega}_{a}=\frac{3}{1/\omega_{\times}+2/\omega_{E_{\nu}}}\,,\label{eq:11}\\
\bar{\lambda}_{g} & =\sqrt[3]{\lambda_{\times}\cdot\lambda_{E_{\nu}}^{2}} & \Longleftrightarrow\quad & \bar{\omega}_{g}=\sqrt[3]{\omega_{\times}\cdot\omega_{E_{\nu}}^{2}}\,,\label{eq:12}\\
\bar{\lambda}_{h} & =\frac{3}{1/\lambda_{\times}+2/\lambda_{E_{\nu}}} & \Longleftrightarrow\quad & \bar{\omega}_{h}=(\omega_{\times}+2\omega_{E_{\nu}})/3\,.\label{eq:13}
\end{align}

The ratios of the above constructed mixed quantities to the dissociation
minimum ($\lambda_{D}$) are shown in Figure \ref{ratio}b. All the
three formulas lead to very similar results predicting the dissociation
minima with less than a half percent error, except for a few cases.
The harmonic mean, defined in eq. (\ref{eq:13}), is the most balanced
combination as its values remain in the narrowest interval around
$\lambda_{D}$. The fact that these wavelengths, especially their
special combination provide an almost perfect molecular stability,
implies that the nodes of the nuclear vibrational wave packet play
an essential role in the bond hardening effect. 

The above numerical results support the theoretical expectations:
initiating the nuclear dissociation dynamics from a given vibrational
level, one can always find the appropriate photon energy that minimizes
the dissociation yield. The latter holds for the total dissociation
probability when the initial wave packet starts from one of the vibrational
eigenstates and for the kinetic energy release spectra (KER) at a
certain $(E_{\nu}+\hbar\omega)$ energy when the initial wave packet
starts from a FC distribution of ionic vibrational eigenstates. At
the heart of this work are the nodes of the nuclear vibrational wave
packet which can quantitatively be associated with the bond hardening
effect.

Our showcase example throughout this study was the $D_{2}^{+}$ molecule.
One may assume that the present findings are independent from $D_{2}^{+}$
and are valid for any other molecular systems, as well.

\section{Conclusions}

In this paper, a series of numerical calculations have been performed,
over a wide range of photon energies in order to provide accurate
results for the vibration trapping phenomenon. It was demonstrated
that the nodal structure of the nuclear vibrational wave packets have
an important role in the molecular stabilization process. We have
found an accurate quantitative connection between the nodes of the
nuclear vibrational wave packet and the minimum of the dissociation
rate. In contrast to the literature, the importance of the nodes of
the vibrational wave packet is strongly stressed because an almost
perfect molecular stabilization can be obtained using our recently
proposed formula (eq. (\ref{eq:13})) of the node based wavelength. 

However, we have yet to clearly identify the background mechanism
behind the harmonic mean combination of the $\lambda_{E_{\nu}}$ and
$\lambda_{\times}$ wavelengths, which can provide the best description
of the phenomena. Our studies will continue and will be extended in
this direction and we expect to form a better understanding about
the complex physical phenomena beyond the vibrational trapping effect.
Nevertheless, we hope that the present findings will stimulate experimental
works in the near future.

\section*{Acknowledgment}

B. P. acknowledges the TÁMOP-4.2.2.B-15/1/KONV-2015-0001 `National
Excellence Program'. Á.V. acknowledges the OTKA (NN103251) project.
The supercomputing service of NIIF has been used for this work. The
ELI-ALPS project (GOP-1.1.1-12/B-2012-000, GINOP-2.3.6-15-2015-00001)
is supported by the European Union and co-financed by the European
Regional Development Fund. The authors thank Lorenz Cederbaum for
many fruitful discussions.

\section*{Author contributions statement }

P.B. performed the numerical simulation. G. J. H and Á. V initiated
and designed the calculations and are responsible for the theoretical
interpretation of the results. P.B and G. J. H. prepared the figures.
Á.V. and G.J.H. wrote the manuscript.

\section*{Additional information}

\textbf{Competing financial interests }

The authors declare no competing financial interests. 

\begin{table}[p]
\centering \global\long\def\arraystretch{1.3}
 {\tiny { }%
\begin{tabular}{lcccccc||lcccccc}
$\nu,\nu'$ & $\lambda_{D}(nm)$ & $\lambda_{O}(nm)$ & $\lambda_{E}(nm)$ & $\lambda_{E_{\nu}}(nm)$ & $\lambda_{\times}(nm)$ & $R_{n}(au)$ & $\nu,\nu'$ & $\lambda_{D}(nm)$ & $\lambda_{O}(nm)$ & $\lambda_{E}(nm)$ & $\lambda_{E_{\nu}}(nm)$ & $\lambda_{\times}(nm)$ & $R_{n}(au)$\tabularnewline
\hline 
{\tiny 1,0 } & {\tiny 111.21 } & {\tiny 123.02 } & {\tiny 120.94 } & {\tiny 112.21 } & {\tiny 109.27 } & {\tiny 2.05 } & {\tiny 7,0 } & {\tiny 258.66 } & {\tiny 291.60 } & {\tiny 281.10 } & {\tiny 266.19 } & {\tiny 243.32 } & {\tiny 3.16 }\tabularnewline
 &  &  &  &  &  &  & {\tiny 7,1 } & {\tiny 206.70 } & {\tiny 222.93 } & {\tiny 219.37 } & {\tiny 215.55 } & {\tiny 189.71 } & {\tiny 2.82 }\tabularnewline
{\tiny 2,0 } & {\tiny 133.27 } & {\tiny 147.85 } & {\tiny 144.81 } & {\tiny 135.15 } & {\tiny 129.57 } & {\tiny 2.29 } & {\tiny 7,2 } & {\tiny 170.00 } & {\tiny 180.76 } & {\tiny 178.66 } & {\tiny 178.28 } & {\tiny 154.89 } & {\tiny 2.54 }\tabularnewline
{\tiny 2,1 } & {\tiny 100.21 } & {\tiny 109.11 } & {\tiny 108.03 } & {\tiny 101.54 } & {\tiny 97.72 } & {\tiny 1.90 } & {\tiny 7,3 } & {\tiny 141.47 } & {\tiny 149.29 } & {\tiny 148.16 } & {\tiny 148.39 } & {\tiny 129.20 } & {\tiny 2.29 }\tabularnewline
 &  &  &  &  &  &  & {\tiny 7,4 } & {\tiny 118.09 } & {\tiny 124.45 } & {\tiny 123.67 } & {\tiny 123.37 } & {\tiny 108.91 } & {\tiny 2.05 }\tabularnewline
{\tiny 3,0 } & {\tiny 154.67 } & {\tiny 171.72 } & {\tiny 167.78 } & {\tiny 157.42 } & {\tiny 149.17 } & {\tiny 2.48 } & {\tiny 7,5 } & {\tiny 98.08 } & {\tiny 103.57 } & {\tiny 103.11 } & {\tiny 101.62 } & {\tiny 91.97 } & {\tiny 1.81 }\tabularnewline
{\tiny 3,1 } & {\tiny 120.08 } & {\tiny 130.09 } & {\tiny 128.75 } & {\tiny 122.67 } & {\tiny 115.18 } & {\tiny 2.13 } & {\tiny 7,6 } & {\tiny 79.78 } & {\tiny 84.98 } & {\tiny 84.69 } & {\tiny 81.55 } & {\tiny 76.74 } & {\tiny 1.56 }\tabularnewline
{\tiny 3,2 } & {\tiny 93.54 } & {\tiny 101.06 } & {\tiny 100.36 } & {\tiny 95.03 } & {\tiny 90.80 } & {\tiny 1.80 } &  &  &  &  &  &  & \tabularnewline
 &  &  &  &  &  &  & {\tiny 8,0 } & {\tiny 293.15 } & {\tiny 331.23 } & {\tiny 319.24 } & {\tiny 302.48 } & {\tiny 274.19 } & {\tiny 3.33 }\tabularnewline
{\tiny 4,0 } & {\tiny 177.09 } & {\tiny 197.19 } & {\tiny 192.21 } & {\tiny 180.79 } & {\tiny 169.62 } & {\tiny 2.66 } & {\tiny 8,1 } & {\tiny 234.31 } & {\tiny 253.22 } & {\tiny 248.51 } & {\tiny 245.42 } & {\tiny 213.03 } & {\tiny 2.98 }\tabularnewline
{\tiny 4,1 } & {\tiny 139.53 } & {\tiny 150.84 } & {\tiny 149.05 } & {\tiny 143.41 } & {\tiny 132.12 } & {\tiny 2.32 } & {\tiny 8,2 } & {\tiny 193.05 } & {\tiny 204.87 } & {\tiny 202.61 } & {\tiny 203.66 } & {\tiny 173.72 } & {\tiny 2.70 }\tabularnewline
{\tiny 4,2 } & {\tiny 111.96 } & {\tiny 120.04 } & {\tiny 119.15 } & {\tiny 114.94 } & {\tiny 106.51 } & {\tiny 2.02 } & {\tiny 8,3 } & {\tiny 161.23 } & {\tiny 169.95 } & {\tiny 168.48 } & {\tiny 170.44 } & {\tiny 145.06 } & {\tiny 2.45 }\tabularnewline
{\tiny 4,3 } & {\tiny 88.78 } & {\tiny 95.42 } & {\tiny 94.97 } & {\tiny 90.38 } & {\tiny 85.91 } & {\tiny 1.72 } & {\tiny 8,4 } & {\tiny 135.49 } & {\tiny 142.27 } & {\tiny 141.41 } & {\tiny 142.92 } & {\tiny 122.74 } & {\tiny 2.21 }\tabularnewline
 &  &  &  &  &  &  & {\tiny 8,5 } & {\tiny 113.90 } & {\tiny 119.54 } & {\tiny 119.00 } & {\tiny 119.44 } & {\tiny 104.50 } & {\tiny 1.99 }\tabularnewline
{\tiny 5,0 } & {\tiny 201.42 } & {\tiny 225.35 } & {\tiny 218.50 } & {\tiny 206.20 } & {\tiny 191.71 } & {\tiny 2.83 } & {\tiny 8,6 } & {\tiny 95.12 } & {\tiny 100.18 } & {\tiny 99.79 } & {\tiny 98.76 } & {\tiny 88.92 } & {\tiny 1.77 }\tabularnewline
{\tiny 5,1 } & {\tiny 159.95 } & {\tiny 172.56 } & {\tiny 170.31 } & {\tiny 165.29 } & {\tiny 149.80 } & {\tiny 2.49 } & {\tiny 8,7 } & {\tiny 77.74 } & {\tiny 82.67 } & {\tiny 82.41 } & {\tiny 79.55 } & {\tiny 74.67 } & {\tiny 1.52 }\tabularnewline
{\tiny 5,2 } & {\tiny 130.10 } & {\tiny 138.91 } & {\tiny 137.62 } & {\tiny 134.62 } & {\tiny 121.78 } & {\tiny 2.20 } &  &  &  &  &  &  & \tabularnewline
{\tiny 5,3 } & {\tiny 106.14 } & {\tiny 113.02 } & {\tiny 112.32 } & {\tiny 109.37 } & {\tiny 100.37 } & {\tiny 1.93 } & {\tiny 9,0 } & {\tiny 332.84 } & {\tiny 377.34 } & {\tiny 363.08 } & {\tiny 344.38 } & {\tiny 309.60 } & {\tiny 3.50 }\tabularnewline
{\tiny 5,4 } & {\tiny 85.13 } & {\tiny 91.25 } & {\tiny 90.69 } & {\tiny 86.79 } & {\tiny 82.17 } & {\tiny 1.66 } & {\tiny 9,1 } & {\tiny 265.72 } & {\tiny 286.96 } & {\tiny 281.69 } & {\tiny 279.52 } & {\tiny 239.35 } & {\tiny 3.14 }\tabularnewline
 &  &  &  &  &  &  & {\tiny 9,2 } & {\tiny 218.95 } & {\tiny 232.59 } & {\tiny 229.52 } & {\tiny 232.31 } & {\tiny 194.68 } & {\tiny 2.85 }\tabularnewline
{\tiny 6,0 } & {\tiny 228.36 } & {\tiny 255.84 } & {\tiny 247.97 } & {\tiny 234.41 } & {\tiny 216.07 } & {\tiny 3.00 } & {\tiny 9,3 } & {\tiny 183.16 } & {\tiny 192.80 } & {\tiny 191.14 } & {\tiny 195.02 } & {\tiny 162.45 } & {\tiny 2.60 }\tabularnewline
{\tiny 6,1 } & {\tiny 182.13 } & {\tiny 196.72 } & {\tiny 193.49 } & {\tiny 189.09 } & {\tiny 168.81 } & {\tiny 2.66 } & {\tiny 9,4 } & {\tiny 154.46 } & {\tiny 162.00 } & {\tiny 160.89 } & {\tiny 164.36 } & {\tiny 137.60 } & {\tiny 2.37 }\tabularnewline
{\tiny 6,2 } & {\tiny 149.22 } & {\tiny 158.75 } & {\tiny 157.26 } & {\tiny 155.50 } & {\tiny 137.74 } & {\tiny 2.37 } & {\tiny 9,5 } & {\tiny 130.69 } & {\tiny 136.82 } & {\tiny 136.08 } & {\tiny 138.46 } & {\tiny 117.57 } & {\tiny 2.15 }\tabularnewline
{\tiny 6,3 } & {\tiny 123.32 } & {\tiny 130.57 } & {\tiny 129.67 } & {\tiny 128.27 } & {\tiny 114.47 } & {\tiny 2.12 } & {\tiny 9,6 } & {\tiny 110.45 } & {\tiny 115.65 } & {\tiny 115.15 } & {\tiny 116.18 } & {\tiny 100.90 } & {\tiny 1.94 }\tabularnewline
{\tiny 6,4 } & {\tiny 101.67 } & {\tiny 107.78 } & {\tiny 107.14 } & {\tiny 105.09 } & {\tiny 95.71 } & {\tiny 1.87 } & {\tiny 9,7 } & {\tiny 92.64 } & {\tiny 97.36 } & {\tiny 97.00 } & {\tiny 96.36 } & {\tiny 86.37 } & {\tiny 1.73 }\tabularnewline
{\tiny 6,5 } & {\tiny 82.20 } & {\tiny 87.80 } & {\tiny 87.40 } & {\tiny 83.92 } & {\tiny 79.19 } & {\tiny 1.61 } & {\tiny 9,8 } & {\tiny 76.00 } & {\tiny 80.70 } & {\tiny 80.47 } & {\tiny 77.82 } & {\tiny 72.91 } & {\tiny 1.49 }\tabularnewline
\hline 
 &  &  &  &  &  &  &  &  &  &  &  &  & \tabularnewline
\end{tabular}{\tiny }}{\small \caption{\label{Table1}{\small Characteristic $\lambda$ wavelengths corresponding
to the $\nu$, $\nu'$ ($\nu'<\nu$) different vibrational diabatic
and adiabatic levels. $R_{n}$ denotes the positions of the nodes
of $\psi_{\nu}$ diabatic wave functions. }}
}
\end{table}

\end{document}